\documentclass{epl}
\usepackage{graphicx}
\usepackage{amsmath}
\usepackage{amsfonts}
\usepackage{amssymb}

\title{Entangling movable mirrors in a double-cavity system}
\author{M. Pinard\inst{1}, A. Dantan\inst{1}, D. Vitali\inst{2}, O. Arcizet\inst{1}, T. Briant\inst{1}, A. Heidmann\inst{1}}
\shortauthor{M. Pinard \textit{et al.}} \institute{ \inst{1}
Laboratoire Kastler Brossel, Universit\'{e} Pierre et
Marie Curie, Case 74, 4 place Jussieu, 75252 Paris Cedex 05, France \\
\inst{2} Dipartimento di Fisica, Universit\`a di Camerino, I-62032
Camerino, Italy } 

\begin{document}

\maketitle

\begin{abstract}
We propose a double-cavity set-up capable of generating a stationary
entangled state of two movable mirrors at cryogenic temperatures.
The scheme is based on the optimal transfer of squeezing of input
optical fields to mechanical vibrational modes of the mirrors,
realized by the radiation pressure of the intracavity light. We show
that the presence of macroscopic entanglement can be demonstrated by
an appropriate read out of the output light of the two cavities.
\end{abstract}

\newcommand{\beq}{\begin{equation}}
\newcommand{\eeq}{\end{equation}}
\newcommand{\beqr}{\begin{eqnarray}}
\newcommand{\eeqr}{\end{eqnarray}}
\newcommand{\lb}[1]{\label{#1}}
\newcommand{\ct}[1]{\cite{#1}}
\newcommand{\dt}{\frac{\partial}{\partial t}}
\newcommand{\dz}{\frac{\partial}{\partial z}}
\newcommand{\nn}{\nonumber}

\section{Introduction} Quantum entanglement is a physical phenomenon in which the quantum
states of two or more systems can only be described with reference
to each other. This leads to correlations between observables of
the systems that cannot be understood on the basis of local
realistic theories \cite{epr,Bell64}. Its importance today exceeds
the realm of the foundations of quantum physics and entanglement
has become an important physical resource that allows performing
communication and computation tasks with an efficiency which is
not achievable classically \cite{Nielsen}. In particular, it is
important to investigate under which conditions entanglement
between macroscopic objects, each containing a large number of the
constituents, can arise. Entanglement between two atomic ensembles
has been successfully demonstrated in Ref.~\cite{juuls01} by
sending pulses of coherent light through two atomic vapor cells.
Then, other proposals suggested to entangle a nano-mechanical
oscillator with a Cooper-pair box \cite{Armour03}, arrays of
nano-mechanical oscillators \cite{eisert04}, two mirrors of an
optical ring cavity \cite{PRL02}, or two mirrors of two different
cavities illuminated with entangled light beams \cite{Peng03}.
Here we elaborate on these two latter schemes and propose a new
double-cavity set-up able to generate a stationary entangled state
of two vibrating cavity mirrors, exploiting the radiation pressure
of the intracavity fields. Entanglement between mechanical degrees
of freedom is achieved if the input fields are squeezed and if
this squeezing is efficiently transferred to the movable mirrors.
We show that a stationary entangled state can be generated with
state-of-the-art apparata at cryogenic temperatures, and that it
can be detected with a non-stationary homodyne measurement of the
output light \cite{dantan1,dantan2}.

\section{The system}
\begin{figure}[htb]
\centerline{\includegraphics[width=6cm,clip=]{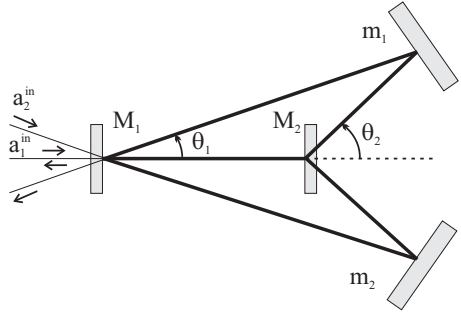}}
\caption{The double-cavity set-up. $M_1$ and $M_2$ are movable
mirrors, while $m_1$ and $m_2$ are fixed. Fields $a_1$, $a_2$ are
sent into the linear and folded cavities, respectively.}
\label{doublecavite}
\end{figure}
Let us consider two movable mirrors $M_1$ and $M_2$, oscillating at
frequency $\Omega_1$ and $\Omega_2$ respectively, which interact
with two field modes $a_1$ and $a_2$, with frequency $\omega_L$
(Fig.~\ref{doublecavite}). Field $a_1$ is injected into a linear
cavity constituted by mirrors $M_1$ and $M_2$, whereas $a_2$ is
injected in a ``folded'' cavity formed by the movable mirrors $M_1$
and $M_2$, and by the fixed mirrors $m_1$ and $m_2$. Field $a_1$ is
coupled to the length of the linear cavity, that is to the relative
position of the mirrors, whereas field $a_2$ is coupled both to the
center-of-mass and relative positions of the mirrors. We will see
that for a judicious choice of the incidence angles $\theta_1$ and
$\theta_2$, field $a_2$ is only coupled to the center of mass
measured with respect to the positions of mirrors $m_1$ and $m_2$.
It is then possible to decouple the relative and center-of-mass
motions of the movable mirrors, as well as their coupling with the
fields.

The motion of the mirrors generally corresponds to a superposition
of many acoustic modes vibrations \cite{Pinard99}. A single
vibrational mode description can however be adopted whenever
detection is limited to a frequency bandwidth including a single
mechanical resonance. Assuming that mirrors $m_1$ and $m_2$ have
much larger masses and no resonance frequency within the detection
bandwidth, their motion can be neglected and the mechanical
hamiltonian of the mirrors is given by \beqr
H_m&=&\frac{P_1^2}{2M_1}+\frac{P_2^2}{2M_2}+
\frac{1}{2}M_1\Omega_1^2Q_1^2+\frac{1}{2}M_2\Omega_2^2Q_2^2.\eeqr
We introduce the operators associated to the relative and the
center of mass motions \beqr \label{cenm}
Q_{cm}&=&\frac{M_1}{M_T}Q_1+\frac{M_2}{M_T}Q_2,\hspace{0.5cm}P_{cm}=P_1+P_2,\\
Q_r&=&Q_1-Q_2,\hspace{1.7cm}\frac{P_r}{\mu}=\frac{P_1}{M_1}-\frac{P_2}{M_2},\label{relm}\eeqr
where $M_T=M_1+M_2$ and $\mu=M_1M_2/M_T$ are the total and reduced
masses of the system, respectively. If both mirrors have equal
resonance frequencies $\Omega_1=\Omega_2=\Omega$, the mechanical
Hamiltonian can be re-expressed as the sum of the Hamiltonians of
two independent harmonic oscillators with frequency $\Omega$ and
masses $\mu$ and $M_T$ \beqr
H_m=\frac{P_{cm}^2}{2M_T}+\frac{1}{2}M_T\Omega^2Q_{cm}^2
+\frac{P_r^2}{2\mu}+\frac{1}{2}\mu \Omega^2Q_r^2.\eeqr The
radiation pressure forces exerted by field $a_1$ on the mirrors
are opposite, whereas they are in the same direction for field
$a_2$. Taking into account the incident angles, the interaction
Hamiltonians resulting from radiation pressure can be written as
\beqr H_1 = \frac{\hbar
\omega_{c1}}{L_1}a_1^{\dagger}a_1(Q_1-Q_2)\hspace{0.5cm}\textrm{and}\hspace{0.5cm}H_2
=\frac{\hbar\omega_{c2}}{L_2}a_2^{\dagger}a_2(Q_1\cos\theta_1+Q_2\cos\theta_2),\eeqr
where $L_1$ and $2L_2$ are the lengths of the linear and folded
cavities respectively, and $\omega_{cj}$ is the frequency of
cavity mode $a_j$. The linear cavity mode $a_1$ is only coupled to
the relative motion, while the folded cavity mode $a_2$ is coupled
both to $Q_r$ and to $Q_{cm}$. One actually finds from
Eqs.~(\ref{cenm},\ref{relm}) \beq\label{eq:H2}
H_2=\frac{\hbar\omega_{c2}}{L_2}a_2^{\dagger}a_2\left(\cos\theta_1+\cos\theta_2\right)Q_{cm}
+\frac{\hbar\omega_{c2}}{L_2}a_2^{\dagger}a_2
\left(-\frac{M_1}{M_T}\cos\theta_1+\frac{M_2}{M_T}\cos\theta_2\right)Q_r.\eeq
If we choose the angles $\theta_1$ and $\theta_2$ so that \beq
\frac{M_1}{M_2}=\frac{\cos\theta_1}{\cos\theta_2},\eeq the second
term in Eq.~(\ref{eq:H2}) vanishes and the radiation pressure
couples $a_2$ to the center-of-mass motion only. We define the
usual annihilation operators $b_r$ and $b_{cm}$ associated to the
relative and center-of-mass motion respectively, as \beqr
&& b_r=\frac{\mu\Omega Q_r+iP_r}{\sqrt{2\hbar\mu\Omega}}=\frac{x_r+ip_r}{\sqrt{2}}, \label{br}\\
&& b_{cm}=\frac{M_T\Omega Q_{cm}+iP_{cm}}{\sqrt{2\hbar
M_T\Omega}}=\frac{x_{cm}+ip_{cm}}{\sqrt{2}},  \label{bcm}\eeqr where
$x_j$ and $p_j$ satisfy $\left[x_j,p_k\right]=i\delta_{jk}$,
($j,k=r,cm$). The radiation pressure interaction terms can then be
rewritten as \beqr H_1=\hbar
G_1a_1^{\dagger}a_1(b_r+b_r^{\dagger})\hspace{1cm}\textrm{and}\hspace{1cm}
H_2=\hbar G_2a_2^{\dagger}a_2 (b_{cm}+b_{cm}^{\dagger}),\eeqr where
$G_1=(\omega_{c1}/L_1)\sqrt{\hbar/2\mu\Omega}$ and
$G_2=(\omega_{c2}/L_2)\sqrt{\hbar/2M_T\Omega}(\cos\theta_1+\cos\theta_2)$
are the optomechanical coupling constants.

The Heisenberg-Langevin equations for the quantum-mechanical
oscillators and the field modes which result from these couplings
are, in the frame rotating at the laser frequency $\omega_L$,
\beqr\label{eq:dota1}
&&\dot{a}_1=-(\kappa_1+i\Delta_1)a_1-iG_1a_1(b_r+b_r^{\dagger})+\sqrt{2\kappa_1}a_1^{in},\\
&&\dot{b}_r=-(\Gamma/2+i\Omega)b_r+(\Gamma/2)b_r^{\dagger}-iG_1a_1^{\dagger}a_1+\xi_r,\label{eq:dotbr}\\
&&\dot{a}_2=-(\kappa_2+i\Delta_2)a_2-iG_2a_2(b_{cm}+b_{cm}^{\dagger})+\sqrt{2\kappa_2}a_2^{in},\label{eq:dota2}\\
&&
\dot{b}_{cm}=-(\Gamma/2+i\Omega)b_{cm}+(\Gamma/2)b_{cm}^{\dagger}-iG_2a_2^{\dagger}a_2+\xi_{cm},\label{eq:dotbcm}
\eeqr where $\kappa_j$ are the cavity bandwidths,
$\Delta_j=\omega_{cj}-\omega_L$ the cavity detunings, and $\Gamma$
the damping rate of the mirrors, assumed equal for both mirrors to
ensure that the center-of-mass and the relative motions are not
coupled via the dissipation process. We thus have two mechanical
oscillators and two optical modes interacting in pairs: the relative
motion only interacts with the linear cavity mode, and the
center-of-mass with the folded cavity mode. The $\delta$-correlated
noise operators $\xi$'s are associated to the Brownian motion of the
mirrors. They have zero-mean value and satisfy \cite{cct} \beqr
\langle
\xi_i(t)\xi_j^{\dagger}(t')\rangle&=&\Gamma(1+n_T)\delta(t-t')\delta_{ij},\label{correcm0}\\
\langle \xi_i^{\dagger}(t)\xi_j(t')\rangle &=&
\Gamma n_T\delta(t-t')\delta_{ij}\label{correcm}
 \eeqr
($i,j=cm,r$), where $n_T=1/(e^{\hbar\Omega/k_BT}-1)$ is the mean
thermal phonon number at equilibrium temperature $T$. This
corresponds to an Ohmic dissipation and the Markovian property is
justified at not too low temperatures, i.e., $\hbar \Omega \ll
kT$.

\section{Steady state and fluctuations} Setting the time-derivatives to zero in the previous equations
yields the steady state values of the intracavity amplitudes and
mirror positions \beqr\nonumber E_1&=&\langle
a_1\rangle=\frac{\sqrt{2\kappa_1}\langle a_1^{in}\rangle
}{\kappa_1+i\Delta_1'},\hspace{0.5cm}\Delta_1'=\Delta_1+G_1\langle
b_r+b_r^{\dagger}\rangle,\hspace{1cm}\langle
b_r\rangle=\frac{-G_1|E_1|^2}{\Omega},\\\nonumber E_2&=&\langle
a_2\rangle=\frac{\sqrt{2\kappa_2}\langle
a_2^{in}\rangle}{\kappa_2+i\Delta_2'},\hspace{0.5cm}\Delta_2'=\Delta_2+G_2\langle
b_{cm}+b_{cm}^{\dagger}\rangle,\hspace{0.5cm}\langle
b_{cm}\rangle=\frac{- G_2|E_2|^2}{\Omega}.\eeqr The effective
detunings $\Delta_j'$ include the mean displacements of the mirrors
due to radiation pressure, which are proportional to the intracavity
intensities $|E_j|^2$. We can arbitrarily choose the detunings
$\Delta_j'$ by setting the detunings of the cavities, as long as we
stay in the stable domain of the bistability induced by the
optomechanical coupling. We consider in the following the case of
detuned cavities with $\Delta_1'=\Delta_2'=\Omega$.

Eqs. (\ref{eq:dota1}-\ref{eq:dotbr}) and
(\ref{eq:dota2}-\ref{eq:dotbcm}) are decoupled and formally
identical: we first treat the case of the relative motion.
Linearizing Eqs.~(\ref{eq:dota1}-\ref{eq:dotbr}) around the steady
state \cite{linearization}, the fluctuations of operators $a_1$
and $b_r$ obey the following equations \beqr\nn &&\delta\dot{
a}_1=-(\kappa_1+i\Delta_1')\delta a_1-iG_1E_1(\delta b_r+\delta
b_r^{\dagger})+\sqrt{2\kappa_1}\delta a_1^{in},\\\nn &&\delta\dot{
b}_r =-(\Gamma/2+i\Omega)\delta b_r+(\Gamma/2)\delta
b_r^{\dagger}-iG_1(E_1^{*}\delta a_1+E_1\delta
a_1^{\dagger})+\xi_r.\eeqr We assume $\Omega\gg\kappa_j$, which
means that the cavities are strongly off-resonant from the fields.
Moving to the frame rotating at frequency $\Omega$ and neglecting
the fast rotating terms, one gets \beqr\label{eq:miroira1tilde} &&
\delta\dot{ \tilde{a}}_1=-\kappa_1\delta
\tilde{a}_1-G_1|E_1|\delta \tilde{b}_r+\sqrt{2\kappa_1}\delta
\tilde{a}_1^{in},\\\label{eq:miroirb1tilde}&& \delta\dot{
\tilde{b}}_r=-(\Gamma/2)\delta \tilde{b}_r+G_1|E_1|\delta
\tilde{a}_1+\tilde{\xi}_r,\eeqr where the slow observables in the
rotating frame are given by $\tilde{o}(t) = o(t)e^{i\Omega t}$ and
we have chosen by convention the phases of the input fields to be
real (so that $E_j=-i|E_j|$). The interesting regime for quantum
state transfer is when the fields adiabatically follow the
mirrors, which is the case for mirrors with high-Q mechanical
factors ($\Gamma, G_j|E_j|\ll\kappa_j$) \cite{briant}.
The mirror dynamics then reduce to \beqr \label{eq:b1mb2}\delta
\dot{\tilde {b}}_r=-\tilde{\gamma}_1\delta
\tilde{b}_r+\sqrt{\Gamma_{c1}}\delta
\tilde{a}_1^{in}+\tilde{\xi}_r,\eeqr where
$\Gamma_{c1}=2G_1^2|E_1|^2/\kappa_1$ represents the effective
relaxation rate induced by radiation pressure \cite{braginsky} and
$\tilde{\gamma}_1=(\Gamma+\Gamma_{c1})/2$. A similar equation can be
obtained from Eqs.~(\ref{eq:dota2},\ref{eq:dotbcm}) for the
center-of-mass operator $b_{cm}$ \beqr\label{eq:b1pb2} \delta
\dot{\tilde {b}}_{cm}=-\tilde{\gamma}_2\delta
\tilde{b}_{cm}+\sqrt{\Gamma_{c2}}\delta
\tilde{a}_2^{in}+\tilde{\xi}_{cm}.\eeqr

\section{Noise spectrum and entanglement} Since $[\tilde{x}_r,\tilde{p}_{cm}]=0$,
the relative position and the center-of-mass momentum of the two
movable mirrors play the role of EPR operators \cite{epr}: their
variances can be simultaneously zero, and if they are small enough,
the corresponding state possesses nonlocal correlations. To be more
specific, according to the inseparability sufficient criterion of
\cite{GIOV03}, the steady state of the two movable mirrors is
entangled if
\begin{equation}  \label{ent2} \Delta x_r^2+\Delta p_{cm}^2
<2\sqrt{\frac{\mu}{M_T}},
\end{equation}
where $\Delta o^2$ denotes the stationary variance of $o$. For
instance, in the case of similar mirrors ($M_1\simeq M_2$), this
criterion is satisfied if the sum of the EPR variances is less than
1.

Assuming for simplicity $\Gamma_{c1}=\Gamma_{c2}=\Gamma_c$, the
Fourier transforms of Eqs.~(\ref{eq:b1mb2}-\ref{eq:b1pb2}) yield
\beqr\label{tildebr}\delta
\tilde{b}_r(\omega)=\frac{\sqrt{\Gamma_c}\delta
\tilde{a}^{in}_1(\omega)+\tilde{\xi}_{r}(\omega)}{\tilde{\gamma}-i\omega},\hspace{0.5cm}\delta
\tilde{b}_{cm}(\omega)&=&\frac{\sqrt{\Gamma_c}\delta
\tilde{a}^{in}_2(\omega)+\tilde{\xi}_{cm}(\omega)}{\tilde{\gamma}-i\omega},\eeqr
with $\tilde{\gamma}=\tilde{\gamma}_1=\tilde{\gamma}_2$. The widths
of the spectra are broadened from $\Gamma/2$ to
$(\Gamma+\Gamma_c)/2$ by radiation pressure effects. The Brownian
motion characterized by the operators $\tilde{\xi}_r$,
$\tilde{\xi}_{cm}$ is reduced by a self-cooling effect for large
$\Gamma_c$ \cite{braginsky}. On the other hand, the quantum
fluctuations of $a_1^{in}$ ($a_2^{in}$) also imprint on $b_r$
($b_{cm}$) when $\Gamma_c\gg\Gamma$. Assuming that fields $a_1^{in}$
and $a_2^{in}$ are respectively amplitude- and phase-squeezed around
$\Omega$ with a bandwidth larger than $\Gamma+\Gamma_c$, the
amplitude fluctuations of $a_1^{in}$ are transferred to $b_r$,
whereas the phase fluctuations of $a_2^{in}$ are transferred to
$p_{cm}$. The EPR variances are then \beqr\label{eq:varxr} \Delta
x_r^2&=&\frac{1}{2}\left[\frac{\Gamma_c}{\Gamma+\Gamma_c}e^{-2r_1}+\frac{\Gamma}{\Gamma+\Gamma_c}(1+2n_T)\right],\\
\Delta
p_{cm}^2&=&\frac{1}{2}\left[\frac{\Gamma_c}{\Gamma+\Gamma_c}e^{-2r_2}+\frac{\Gamma}{\Gamma+\Gamma_c}(1+2n_T)\right],\eeqr
where $\Delta^2 X_1^{in}=e^{-2r_1}$, $\Delta^2 Y_2^{in}=e^{-2r_2}$
are the variances of the amplitude and phase quadratures,
respectively [$X=a+a^{\dagger}$ and $Y=i(a^{\dagger}-a)$]. These
equations stress the physical processes able to produce the
stationary entanglement: first, the mirrors thermal noise is reduced
by self-cooling when $\Gamma_c\gg 2\Gamma n_T$. Secondly, the
quantum fluctuations of the input fields are transferred to the
center-of-mass and relative motions of the two mirrors with an
optimal efficiency at the resonance condition
$\Delta_1'=\Delta_2'=\Omega$. This process is very similar to the
quantum state transfer from light to atomic variables
\cite{dantan1,dantan2}.

One gets for coherent inputs ($r_1=r_2=0$) \beqr\nn \Delta
x_r^2=\Delta p_{cm}^2=\frac{1}{2}+\frac{\Gamma}{\Gamma+\Gamma_c} n_T.
\eeqr The relative position and the center-of-mass momentum can
never be squeezed, and therefore the two mirrors are never
entangled. When the incident fields are squeezed, it is clear that
the mirror motions reproduce the squeezed fluctuations of the
incident fields for a large enough cooling rate $ \Gamma_c/\Gamma
\gg 2n_T$. At a given temperature, this condition can always be
achieved for large enough intracavity intensities. For almost
identical masses ($M_1\simeq M_2$), both mirrors are then in an
entangled state: \beqr \Delta x_r^2+\Delta p_{cm}^2\simeq
\frac{1}{2}(e^{-2r_1}+e^{-2r_2})<1.\eeqr For perfectly squeezed
inputs, this corresponds to a realization of the original EPR
paradox for the mechanical oscillators, with perfect position
correlations and momentum anti-correlations.

\begin{figure}[htb]
\centerline{\includegraphics[width=9cm]{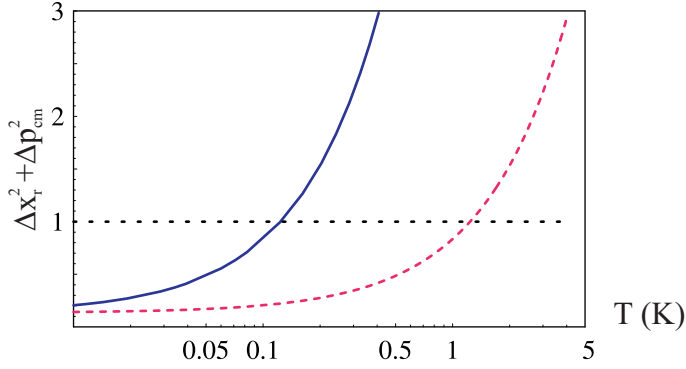}} \caption{Plots of
the normalized sum of variances $\Delta x_r^2+ \Delta p_{cm}^2$
versus temperature, in the case of almost identical mirrors
$M_1\simeq M_2=1$ mg (full line), $M_1\simeq M_2=1$ $\mu$g (dashed
line). The dotted horizontal line denotes the region below which the
stationary state of the two movable mirrors is entangled. See text
for the parameter values.} \label{entprod-eqfreq}
\end{figure}

Such a stationary entanglement can already be achieved with
state-of-the-art apparata, working at cryogenic temperatures, as
shown in Fig.~\ref{entprod-eqfreq}, where the (normalized)
criterion (\ref{ent2}) is plotted versus the mirror temperature.
The first curve refers to two identical mirrors with $M_1\simeq
M_2=1$ mg, $\Omega/2\pi = 1$ MHz, $\Gamma=1$ Hz, cavities with
finesse $10^5$, lengths $L_1=3$ cm, $L_2=9$ cm, input squeezing
such that $r_1=r_2=2$, and input powers $P_1^{in}=30$ mW,
$P_2^{in}=1.2$ W ($\Gamma_c/\Gamma=10^3$). The second curve
corresponds to two micro-electro mechanical mirrors (MEMS) with
masses 1 $\mu$g, for which much smaller input powers
$P_1^{in}=0.3$ mW, $P_2^{in}=12$ mW ($\Gamma_c/\Gamma=10^4)$ and
not as low temperatures are required. Thanks to the self-cooling
process the quantum state transfer is efficient even at relatively
high temperatures.

\section{Readout} The motion of a movable mirror is usually
measured by monitoring the phase of the field reflected by a
high-finesse resonant cavity \cite{PRL02,hadjar}. In this paper,
we adopt another strategy to readout the quantum noise of the
mirrors, inspired by optical readout of atomic ensemble quantum
states \cite{dantan1,dantan2}. Let us assume that after completion
of the fluctuation transfer between the fields and the mirrors,
one rapidly switches off the squeezings entering the cavities, the
field intensities being kept constant. We therefore start in a
regime in which the thermal noise has been damped by the self
cooling process and we want to read out the mirror squeezed
fluctuations of the fields exiting the cavities:
$a_j^{out}=\sqrt{2\kappa_j}a_j-a_j^{in}$.

Denoting by $t=0$ this switching time, one gets after integrating
(\ref{eq:b1mb2}) \beq\label{eq:correlation} \delta
\tilde{a}_1^{out}(t)=\delta
\tilde{a}_1^{in}(t)-\Gamma_c\int_0^tdt'e^{-\tilde{\gamma}(t-t')}\delta
\tilde{a}_1^{in}(t')-\sqrt{\Gamma_c}\delta
\tilde{b}_r(0)e^{-\tilde{\gamma}t}-\sqrt{\Gamma_c}\int_0^tdt'e^{-\tilde{\gamma}(t-t')}\tilde{\xi}_{r}(t').\eeq
The term proportional to $\delta \tilde{b}_r(0)$ carries the
information to be measured. The two-time correlation function of the
outgoing amplitude quadrature is then given by \beq\label{eq:correl}
\langle\delta \tilde{X}_1^{out}(t)\delta
\tilde{X}_1^{out}(t')\rangle
=\delta(t-t')+\frac{n_T\Gamma\Gamma_c}{\tilde{\gamma}}\;e^{-\tilde{\gamma}|t-t'|}
+2\Gamma_c\left[\Delta
x_r^2(0)-\frac{\Gamma_c+\Gamma(1+2n_T)}{2(\Gamma+\Gamma_c)}\right]\;e^{-\tilde{\gamma}(t+t')}.\eeq
The two first terms give the $\delta$-correlated function of the
field in the absence of coupling and the contribution of the mirror
thermal noise, respectively. The last term is proportional to the
difference between the initial variance $\Delta x_r^2(0)$ and the
final one $\Delta x_r^2(\infty)$, corresponding to the thermal
equilibrium in the self-cooled regime (see Eq. (\ref{eq:varxr}) with
$r_1=0$).

In order to efficiently measure the stored squeezing, we perform a
homodyne detection of the outgoing field fluctuations using a
local oscillator with a temporal profile matching that of the
mirror response: $E(\tau)\varpropto e^{-\tilde{\gamma}\tau}$. We
then measure the noise starting at a given time $t$ and
integrating over a time $t_m$ assumed large with respect to
$\tilde{\gamma}^{-1}$. The noise spectrum is integrated around
frequency $\Omega$ with a frequency bandwidth $\Delta
\omega=2\pi/t_m$. The resulting normalized noise power is equal to
\beqr
P_1(t)=\frac{1}{I(t)}\int_{-\Delta\omega/2}^{\Delta\omega/2}\frac{d\omega}
{2\pi}\int_t^{t+t_m}d\tau\int_t^{t+t_m}d\tau'
E(\tau)E^*(\tau')e^{-i\omega(\tau-\tau')}\langle\delta
\tilde{X}_1^{out}(\tau)\delta \tilde{X}_1^{out}(\tau')\rangle,
\eeqr where $I(t)=E(t)E^*(t)$. From Eq. (\ref{eq:correl}) this
noise power can be written as the sum of constant noise terms and
a signal term depending on the initial time $t$, \beqr
P_1(t)&=&\frac{1}{2\tilde{\gamma}t_m}\left[1+\frac{4 \Gamma
\Gamma_c}{\left(\Gamma+\Gamma_c\right)^2} n_T +
\frac{2\Gamma_c}{\Gamma+\Gamma_c}\left[\Delta x_r^2(0)-\Delta
x_r^2(\infty)\right]e^{-2\tilde{\gamma}t}\right].\eeqr The first
two terms represent the field shot-noise level and the
contribution of the mirror thermal noise, whereas the signal term
is proportional to the difference between the mirror squeezed
initial variance and its variance at thermal equilibrium. If the
self-cooling is strong enough to reduce the thermal noise
($\Gamma_c/\Gamma\gg 1,2n_T$), the noise terms reduce to 1 and
$\Delta x_r^2(\infty)\sim 1/2$, so that one effectively measures
squeezing in the outgoing field, leaking out of the cavity in a
time $\tilde{\gamma}^{-1}$ and directly related to the squeezing
$\left(1/2-\Delta x_r^2(0)\right)$ initially stored in the
mirrors. Of course, measuring the fluctuations of
$\tilde{Y}_2^{out}$ would yield a similar measurement of $\Delta
p_{cm}^2(0)$. This readout technique thus provides an unambiguous
and experimentally accessible evidence that the mirrors were
entangled. Indeed, if one were to use motionless mirrors for
instance, the squeezing exiting the cavity would disappear in a
time $t\sim\kappa^{-1}$ much shorter than $\tilde{\gamma}^{-1}$.

\section{Conclusion} We have proposed a new scheme for generating
and detecting a stationary entangled state of two vibrational modes
of a pair of mirrors in a double-cavity system. Entanglement is
achievable at few Kelvin degrees if the input light is appropriately
squeezed and for sufficiently large intracavity light intensities
and small mirrors (MEMS).

This work was partially supported by the COVAQIAL European Project
No. FP6-511004. Laboratoire Kastler Brossel is an Unit\'e Mixte du
Centre National de la Recherche Scientifique, de l'Ecole Normale
Sup\'erieure et de l'Universit\'e Pierre et Marie Curie.

\end{document}